\documentclass[12pt]{article}
\usepackage{epsfig}
\usepackage{psfrag}
\usepackage{latexsym}
\def\beq{\begin{equation}}
\def\eeq{\end{equation}}
\def\bea{\begin{eqnarray}}
\def\eea{\end{eqnarray}}
\def\bet{\begin{tabular}}
\def\eet{\end{tabular}}
\def\bes{\begin{subequations}\bea}
\def\ees{\eea\end{subequations}}

\def\be{\begin{equation}}
\def\ee{\end{equation}}
\def\bc{\begin{center}}
\def\ec{\end{center}}
\def\bea{\begin{eqnarray}}
\def\eea{\end{eqnarray}}
\def\dd{\displaystyle}
\def\nn{\nonumber}

\catcode`@=11
\def\marginnote#1{}
\newcount\hour
\newcount\minute
\newtoks\amorpm
\hour=\time\divide\hour by60
\minute=\time{\multiply\hour by60 \global\advance\minute by-\hour}
\edef\standardtime{{\ifnum\hour<12 \global\amorpm={am}%
        \else\global\amorpm={pm}\advance\hour by-12 \fi
        \ifnum\hour=0 \hour=12 \fi
        \number\hour:\ifnum\minute<10 0\fi\number\minute\the\amorpm}}
\edef\militarytime{\number\hour:\ifnum\minute<10 0\fi\number\minute}
\def\draftlabel#1{{\@bsphack\if@filesw {\let\thepage\relax
   \xdef\@gtempa{\write\@auxout{\string
      \newlabel{#1}{{\@currentlabel}{\thepage}}}}}\@gtempa
   \if@nobreak \ifvmode\nobreak\fi\fi\fi\@esphack}
        \gdef\@eqnlabel{#1}}
\def\@eqnlabel{}
\def\@vacuum{}
\def\draftmarginnote#1{\marginpar{\raggedright\scriptsize\tt#1}}
\def\draft{\oddsidemargin 0.0truein
        \def\@oddfoot{\sl preliminary draft \hfil
        \rm\thepage\hfil\sl\today\quad\militarytime}
        \let\@evenfoot\@oddfoot \overfullrule 3pt
        \let\label=\draftlabel
        \let\marginnote=\draftmarginnote
   \def\@eqnnum{(\theequation)\rlap{\kern\marginparsep\tt\@eqnlabel}%
\global\let\@eqnlabel\@vacuum}  }
\catcode`@=12
\begin{document}
\begin{titlepage}
\vspace*{-1cm}
\phantom{hep-ph/***} 
\hfill{DFPD-06/TH/12}

\hfill{RM3-TH/06-17}

\hfill{CERN-PH-TH/2006-205}
\vskip 2.5cm
\begin{center}
{\Large\bf Tri-bimaximal Neutrino Mixing from Orbifolding}
\end{center}
\vskip 0.2  cm
\vskip 0.5  cm
\begin{center}
{\large Guido Altarelli}~\footnote{e-mail address: guido.altarelli@cern.ch}
\\
\vskip .1cm
Dipartimento di Fisica `E.~Amaldi', Universit\`a di Roma Tre
\\ 
INFN, Sezione di Roma Tre, I-00146 Rome, Italy
\\
\vskip .1cm
and
\\
CERN, Department of Physics, Theory Division
\\ 
CH-1211 Geneva 23, Switzerland
\\

\vskip .2cm
{\large Ferruccio Feruglio}~\footnote{e-mail address: feruglio@pd.infn.it} and 
{\large Yin Lin}~\footnote{e-mail address: Lin@pd.infn.it}
\\
\vskip .1cm
Dipartimento di Fisica `G.~Galilei', Universit\`a di Padova 
\\ 
INFN, Sezione di Padova, Via Marzolo~8, I-35131 Padua, Italy
\\
\end{center}
\vskip 0.7cm
\begin{abstract}
\noindent

We show that the $A_4$ discrete symmetry that naturally leads to tri-bimaximal 
neutrino mixing can be simply obtained as a result of  
an orbifolding starting from a model in 6 dimensions. This particular orbifolding has four fixed points where 4 dimensional branes  are located
and the tetrahedral symmetry  
of $A_4$ connects these branes. In this approach $A_4$ appears after 
the reduction from six to four dimensions as a remnant of the 6D 
space-time symmetry.
A previously discussed supersymmetric  
version of $A_4$ is reinterpreted along these lines.

\end{abstract}
\end{titlepage}
\setcounter{footnote}{0}
\vskip2truecm

\section{Introduction}
It is an experimental fact \cite{data} that within measurement errors  
the observed neutrino mixing matrix is compatible with  
the so called tri-bimaximal form, introduced by Harrison, Perkins  
and Scott (HPS) \cite{hps}. It is an interesting challenge to  
formulate dynamical principles that, in a completely natural way, can  
lead to this specific mixing pattern as a first approximation, with  
small corrections determined by higher order terms in a well defined  
expansion. In a series of papers \cite{ma1,ma2} it has been  
pointed out that a broken flavour symmetry based on the discrete  
group $A_4$ appears to be particularly fit for this purpose. Other  
solutions based on continuous flavour groups like SU(3) or SO(3) have  
also been recently presented \cite{continuous,others}, but the $A_4$  
models have a very economical   
and attractive structure (for example, in terms of field content). 
A crucial feature of all HPS models is the  
mechanism used to guarantee the necessary VEV alignment of the flavon  
field $\varphi_T$ which determines the charged lepton mass matrix  
with respect to the
direction in flavour space chosen by the flavon $\varphi_S$ that  
gives the neutrino mass matrix.
In  recent papers \cite{OurTriBi,Mod} we have constructed   
explicit versions of $A_4$ model where the alignment problem is  
solved.  In ref. \cite{OurTriBi} we adopted an extra dimensional  
framework, with $\varphi_T$ and $\varphi_S$ on different branes so  
that the minimization of the respective potentials is kept to a large  
extent independent. 
In ref. \cite{Mod}, we presented an alternative,  
perhaps more conventional, formulation of the $A_4$ model in 4  
dimensions with supersymmetry (SUSY) at the price of introducing a  
somewhat  less economic set of fields. Versions either with see-saw  
or without see-saw can be constructed. The existence of different  
realizations shows that the connection of $A_4$ with the HPS matrix  
is robust and does not necessarily require extra dimensions.

Another important aspect of the problem is that of trying to  
understand the dynamical origin of $A_4$. As a first move in this  
direction, in ref. \cite{Mod} we have reformulated  $A_4$  as a  
subgroup of the modular group which often plays a  
role in the formalism of string theories, for example in the context  
of duality transformations \cite{gpr}.
In the present note we show that the $A_4$ symmetry can be simply  
obtained by orbifolding starting with a model in 6 dimensions (6D).  
In this approach $A_4$ appears as the remnant of the reduction
from 6D to 4D space-time symmetry induced by the 
special orbifolding adopted.  
There are 4D branes at the four fixed points of the orbifolding and the  
tetrahedral symmetry of $A_4$ connects these branes. The standard  
model fields have components on the fixed point branes while the scalar  
fields necessary for the $A_4$ breaking are in the bulk.

In this paper, starting from a 6D field theory, we first introduce  
the specific orbifolding with four fixed points on which the 4D  
standard model fields live (while a number of additional gauge  
singlets are in the bulk) and specify how the $A_4$ transformations  
relate the field components on different branes or on the bulk. We  
then study the invariant interactions, local in 6D, constructed out  
of the fields in the theory which are invariant under $A_4$. Finally  
we rederive the SUSY model for tri-bimaximal neutrino mixing  
in this particular framework.
%
%
\section{$A_4$ as the isometry of $T^2/Z_2$}

We consider a quantum field theory in 6 dimensions, with two extra dimensions
compactified on an orbifold $T^2/Z_2$. We denote by $z=x_5+i x_6$ the complex
coordinate describing the extra space. The torus $T^2$ is defined by identifying 
in the complex plane the points related by
\be
\begin{array}{l}
z\to z+1\\
z\to z+\gamma~~~~~~~~~~~~~~~~~\gamma=e^{\dd i\frac{\pi}{3}}~~~,
\label{torus}
\end{array}
\ee
where our length unit, $2\pi R$, has been set to 1 for the time being.
The parity $Z_2$ is defined by
\be
z\to -z
\label{parity}
\ee
and the orbifold $T^2/Z_2$ can be represented by the fundamental region given by the triangle
with vertices $0,1,\gamma$, see Fig. 1. The orbifold has four fixed points, $(z_1,z_2,z_3,z_4)=(1/2,(1+\gamma)/2,\gamma/2,0)$.
The fixed point $z_4$ is also represented by the vertices $1$ and $\gamma$. In the orbifold,
the segments labelled by $a$ in Fig. 1, $(0,1/2)$ and $(1,1/2)$, are 
identified and similarly for those labelled by $b$, $(1,(1+\gamma)/2)$ and 
$(\gamma,(1+\gamma)/2)$, and those labelled by $c$, $(0,\gamma/2)$, $(\gamma,\gamma/2)$. Therefore the orbifold is a regular tetrahedron
with vertices at the four fixed points.
\begin{figure}[h!]
\centerline{\psfig{file=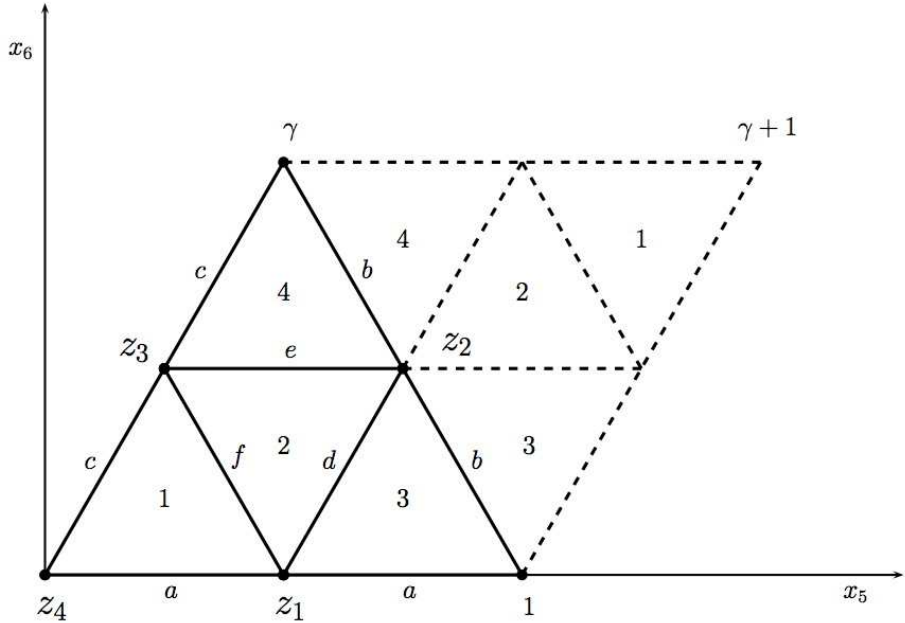,width=0.8\textwidth}}
\caption{Orbifold $T_2/Z_2$. The regions with the same numbers are 
identified with each other. The four triangles bounded by solid lines form the
fundamental region, where also the edges with the same letters are identified.
The orbifold $T_2/Z_2$ is exactly a regular tetrahedron with 6 edges
$a,b,c,d,e,f$ and four vertices $z_1$, $z_2$, $z_3$, $z_4$, corresponding to 
the four fixed points of the orbifold. }
\end{figure}
The symmetry of the uncompactified 6D space time is broken
by compactification. Here we assume that, before compactification,
the space-time symmetry coincides with the product of 6D translations
and 6D proper Lorentz transformations. The compactification breaks
part of this symmetry.
However, due to the special geometry of our orbifold, 
a discrete subgroup of rotations and translations in the extra space is left
unbroken. This group can be generated by two transformations:
\be
\begin{array}{ll}
{\cal S}:& z\to z+\frac{1}{2}\\
{\cal T}:& z\to \omega z~~~~~~~~~~~~~\omega\equiv\gamma^2~~~~.
\label{rototra}
\end{array}
\ee
Indeed ${\cal S}$ and ${\cal T}$ induce even permutations of the four fixed points:
\be
\begin{array}{cc}
{\cal S}:& (z_1,z_2,z_3,z_4)\to (z_4,z_3,z_2,z_1)\\
{\cal T}:& (z_1,z_2,z_3,z_4)\to (z_2,z_3,z_1,z_4)
\end{array}~~~,
\label{stfix}
\ee
thus generating the group $A_4$. \footnote{Notice that an odd permutation of
the four fixed points can be generated by the parity:
\be
z\to z^*~~~,
\label{zparity}
\ee
that maps $(z_1,z_2,z_3,z_4)$ into $(z_1,z_3,z_2,z_4)$ and belongs to the full
6D Poincar\'e group, which, beyond 6D translations and proper Lorentz
transformations, also includes discrete symmetries. Therefore, 
had we assumed 6D Poincar\'e as starting point in the uncompactified theory,
we would have ended up with 
the product of 4D Poincar\'e times the discrete group $S_4$.} From 
the previous equations we immediately verify that ${\cal S}$ and ${\cal T}$ satisfy
the characteristic relations obeyed by the generators of $A_4$:
\be
{\cal S}^2={\cal T}^3=({\cal ST})^3=1~~~.
\label{relations}
\ee
These relations are actually satisfied not only at the fixed points, but on the whole orbifold,
as can be easily checked from the general definitions of ${\cal S}$ and ${\cal T}$ in eq. (\ref{rototra}),
with the help of the orbifold defining rules in eqs. (\ref{torus}) and (\ref{parity}).
In our model the discrete group $A_4$, together with 4D translations and 4D proper Lorentz transformations, 
can be seen as the subgroup of the space-time symmetry in six dimensions that survives
compactification. In a similar context, the compactification of two extra 
dimensions on an orbifold $T^2/Z_3$ and its relation to the flavour symmetry $Z_3$ has been analyzed in ref. (\cite{watyan}).

It is useful to represent the action of ${\cal S}$ and ${\cal T}$ on the fixed points
by means of the four by four matrices $S$ and $T^{-1}$ respectively.
\be
S=
\left(
\begin{array}{cccc}
0& 0& 0& 1\\
0& 0& 1& 0\\
0& 1& 0& 0\\
1& 0& 0& 0
\end{array}
\right)~~~,~~~~~~~~~~
T^{-1}=
\left(
\begin{array}{cccc}
0& 1& 0& 0\\
0& 0& 1& 0\\
1& 0& 0& 0\\
0& 0& 0& 1
\end{array}
\right)~~~.
\label{st4}
\ee
The matrices $S$ and $T$ satisfy the relations (\ref{relations}), thus providing a representation
of $A_4$. Since the only irreducible representations of $A_4$ are a triplet and three singlets,
the 4D representation described by $S$ and $T$ is not irreducible. It decomposes
into the sum of the invariant singlet plus the triplet representation. 
If we denote by $u=(u_1,u_2,u_3,u_4)^t$ (the suffix $t$ denotes transposition) a multiplet transforming as
\be
u\to S u~~~,~~~~~~~u\to T u~~~,
\label{utrans}
\ee
under $S$ and $T$ respectively, then singlet corresponds to
\be
u_1=u_2=u_3=u_4~~~,
\label{sing}
\ee
while the triplet is obtained by imposing the constraint
\be
\sum_{i=1}^{4} u_i=0~~~.
\label{trip}
\ee
Both conditions (\ref{sing}) and (\ref{trip}) are invariant under $A_4$.
To better visualize this decomposition, we consider the unitary matrix
$U$ given by:
\be
U=\frac{1}{2}
\left(
\begin{array}{cccc}
+1&+1&+1&+1\\
& & &\\
-1&+1&+1&-1\\
& & &\\
+1&-1&+1&-1\\
& & &\\
+1&+1&-1&-1
\end{array}
\right)~~~.
\label{U}
\ee
This matrix maps $S$ and $T$ into matrices that 
are block-diagonal:
\be
U S U^\dagger=
\left(
\begin{array}{c|ccc}
1& &0&\\
\hline
& & &\\
0& &S_3&\\
& & &
\end{array}
\right)~~~,~~~~~~~~~~
U T U^\dagger=
\left(
\begin{array}{c|ccc}
1& &0&\\
\hline
& & &\\
0& &T_3&\\
& & &
\end{array}
\right)~~~,
\ee
where $S_3$ and $T_3$ are the generators of the three-dimensional representation:
\be
S_3=
\left(
\begin{array}{ccc}
1& 0& 0\\
0& -1& 0\\
0& 0& -1
\end{array}
\right)~~~,~~~~~~~~~~
T_3=
\left(
\begin{array}{ccc}
0& 0& 1\\
1& 0& 0\\
0& 1& 0
\end{array}
\right)~~~.
\ee
If $u=(u_1,u_2,u_3,u_4)^t$ transforms as specified in eq. (\ref{utrans}),
then $v\equiv(v_0,v_1,v_2,v_3)^t=U u$ transforms as
\be
v\to (U S U^\dagger) v~~~,~~~~~~~v\to (U T U^\dagger) v~~~,
\ee
respectively. Therefore, if we parametrize $u$ as
\be
\left(
\begin{array}{c}
u_1\\
u_2\\
u_3\\
u_4
\end{array}
\right)=\dd\frac{1}{2}
\left(
\begin{array}{c}
v_0\\
v_0\\
v_0\\
v_0
\end{array}
\right)
+\dd\frac{1}{2}
\left(
\begin{array}{c}
-v_1+v_2+v_3\\
+v_1-v_2+v_3\\
+v_1+v_2-v_3\\
-v_1-v_2-v_3
\end{array}
\right)~~~,
\label{decomp}
\ee
the components $(v_1,v_2,v_3)^t$ transform with $S_3$ and $T_3$, whereas the component $v_0$
is left invariant by $A_4$. It is useful to observe that $v_0$ is given by 
$v_0=(u_1+u_2+u_3+u_4)/2$ while
the sum of all components of the last multiplet in eq. (\ref{decomp}) vanishes,
in agreement with the conditions (\ref{sing}) and (\ref{trip}).
Finally, if we restrict to the case of a pure triplet by taking $v_0=0$, then 
$v_1$, $v_2$ and $v_3$ are given by:
\be
\left(
\begin{array}{c}
0\\
v_1\\
v_2\\
v_3
\end{array}
\right)=
U
\left(
\begin{array}{c}
u_1\\
u_2\\
u_3\\
-u_1-u_2-u_3
\end{array}
\right)=
\left(
\begin{array}{c}
0\\
u_2+u_3\\
u_1+u_3\\
u_1+u_2
\end{array}
\right)~~~.
\ee

%
\section{Local interactions invariant under $A_4$}
In this section we collect the rules to construct an $A_4$ invariant field theory
in the 6D space-time ${\cal M}\times T^2/Z_2$.
The fields of this theory can be either 4D fields living at the fixed points,
in short `brane' fields, or `bulk' fields depending on both the uncompactified  coordinates $x$ 
and the complex coordinate $z$. The new essential feature with respect to
a 4D formalism is that in general all particles have components over all
four fixed points.
Locality in 6D implies that at each fixed point only products of components on that brane
are allowed in the interaction terms. This constraint reduces the number of invariant interactions
that can be constructed out of brane fields. We now discuss the structure of the invariants
in this context.

\subsection{Brane fields}

We first consider the case of brane fields and we denote by
\be
a=(a_1(x),a_2(x),a_3(x),a_4(x))
\ee
a set of fields localized at the fixed points $(z_1,z_2,z_3,z_4)$, respectively.
For the time being we do not specify if $a$ is a scalar, a spinor or a vector under the 4D 
Lorentz group.
We denote by $\delta_i=\delta(z-z_i)$ the 2D Dirac deltas needed
to construct an interaction term local in 6D, starting from brane fields.
We observe that, if $z$ undergoes the transformations
(\ref{rototra}), then the delta functions $\delta=(\delta_1,\delta_2,\delta_3,\delta_4)^t$ 
are mapped into \footnote{Notice that the action of ${\cal T}$ on the Dirac deltas is described by $T$, the inverse of the matrix $T^{-1}$
that permutes the four fixed points, eq. (\ref{st4}).}
\be
\begin{array}{cc}
{\cal S}:& \delta\to S\delta\\
{\cal T}:& \delta\to T\delta
\end{array}~~~,
\label{stdelta}
\ee
where $S$ and $T$ are given in eq. (\ref{st4}).
The $A_4$ transformations of $a$ are naturally given by:
\be
\begin{array}{cl}
{\cal S}:& a\to S a\\
{\cal T}:& a\to T a
\end{array}~~~,
\label{a4brane}
\ee
According to our discussion in the previous section, the quadruplet $a$ decomposes into a triplet plus the invariant singlet 1.
If we introduce two such sets of brane fields, called $a$ and $b$, transforming as specified in eq. (\ref{a4brane}),
then it is easy to see that the only invariant under $A_4$, bilinear in $a$ and $b$ and
local in 6D is given by:  
\be
J^{(2)}=\sum_{i=1}^{4} a_i b_i \delta_i~~~.
\label{2inv}
\ee
In particular, if $a=(a_c/2,a_c/2,a_c/2,a_c/2)$ and $b=(b_c/2,b_c/2,b_c/2,b_c/2)$ are two invariant singlets, then, after integrating over the $z$ coordinate, 
the invariant $J^{(2)}$ is given by $\int d^2z J^{(2)}=a_c b_c$. 
If $a$ is a singlet and $b$ is a triplet, $J^{(2)}$ 
vanishes after integration
over $z$, because of eq. (\ref{trip}). If $a$ and $b$ are two triplets transforming as in eq. (\ref{a4brane}), they can be parametrized as
shown in eq. (\ref{decomp}):
\be
a=\frac{1}{2}
\left(
\begin{array}{c}
-v_1+v_2+v_3\\
+v_1-v_2+v_3\\
+v_1+v_2-v_3\\
-v_1-v_2-v_3
\end{array}
\right)~~~,~~~~~~~
b=\frac{1}{2}
\left(
\begin{array}{c}
-w_1+w_2+w_3\\
+w_1-w_2+w_3\\
+w_1+w_2-w_3\\
-w_1-w_2-w_3
\end{array}
\right)~~~.
\label{parab}
\ee
In this case, after integration over $z$, the bilinear $J^{(2)}$ reads:
\be
\int d^2z J^{(2)}=v_1 w_1+v_2 w_2+v_3 w_3~~~,
\ee
which is the familiar expression of the invariant under $A_4$ contained in the product
of two triplet representations \cite{OurTriBi}.

Locality in 6D provides some limitations in the construction of interaction terms.
For instance, it will be important for the following discussion to note that 
if $a$ and $b$ are two triplets transforming as in (\ref{a4brane}), then it is not possible
to construct a term bilinear in $a$ and $b$, local in 6D and transforming as a $1'$
or a $1''$. This is easily seen by starting from the local bilinear
\be
J'=\sum_{i=1}^4 y_i a_i b_i \delta_i~~~,
\ee
where $y_i$ are constants to be determined by imposing that $J'$ transforms as a $1'$.
In fact it is trivial to see that only the trivial solution $y_i=0$ is allowed.
This is because ${\cal S}$ imposes $y_4=y_1$ and $y_3=y_2$; while ${\cal T}$ requires
$y_4=\omega y_4$, hence $y_4=y_1=0$, and $y_1=\omega y_2=\omega^2 y_2$, so that also
$y_2=y_3=0$. The same argument also shows that it is equally impossible to obtain $1''$.

To obtain a non-invariant singlet from two triplets one has two possibilities. The first one is to exploit
bulk fields, as we shall see in detail in the next subsection. The second one is to make use of a freedom
associated to the $A_4$ algebra, by generalizing the transformation properties of the brane fields
in the following way:
\be
\begin{array}{cl}
{\cal S}:& a\to S a\\
{\cal T}:& a\to \omega^{r_a} T a
\end{array}~~~,
\label{a4branenew}
\ee
where $\omega$ is a cubic root of unity, eq. (\ref{rototra}), and $r_a=(0,\pm 1)$. 

Clearly these new transformations
satisfy the $A_4$ algebra, eq. (\ref{relations}). The only difference with respect to the transformations
in eq. (\ref{a4brane}) is in the phase factor $\omega^{r_a}$. It is possible to show that, once the delta function
transformations are specified as in eq. (\ref{stdelta}), then eq. (\ref{a4branenew}) provides the  
only allowed generalization of eq. (\ref{a4brane}).
If we call ${\cal R}_{0,-1,+1}$ these
representations, we see that they are all reducible: ${\cal R}_0$ decomposes
into a triplet plus the invariant singlet 1, ${\cal R}_{+1}$ decomposes
into a triplet plus the singlet $1'$ and ${\cal R}_{-1}$ decomposes
into a triplet plus the singlet $1''$. 
It is immediate to see that $J^{(2)}$ is left invariant by $A_4$ only if $(a,b)$ transform as ${\cal R}_a,{\cal R}_b$
with $a+b=0$.
To build a non-invariant singlet one has to assign $(a,b)$ to 
$({\cal R}_0,{\cal R}_{\pm 1})$. For example, consider the case ${\cal R}=+1$ for $b$. 
Then the triplet $(w_1,w_2,w_3)$ can be embedded in $b$ in the following way:
\be
b=\frac{1}{2}
\left(
\begin{array}{c}
-w_1+\omega w_2+\omega^2 w_3\\
+w_1-\omega w_2+\omega^2 w_3\\
+w_1+\omega w_2-\omega^2 w_3\\
-w_1-\omega w_2-\omega^2 w_3
\end{array}
\right)~~~.
\label{parb}
\ee
Now the bilinear
\be
\sum_{i=1}^{4} a_i b_i \delta_i~~~,
\label{2singl}
\ee
is invariant under ${\cal S}$ and picks up a phase $\omega$ under ${\cal T}$, that is
it transforms as a singlet $1'$. After integrating over
the coordinate $z$, we find
\be
\int d^2z \sum_{i=1}^{4} a_i b_i \delta_i=v_1 w_1+\omega v_2 w_2+\omega^2 v_3 w_3~~~.
\label{2singl2}
\ee
This example shows that, although from the point of view of the group $A_4$ the
triplet representations contained in ${\cal R}_0$, ${\cal R}_{+1}$, ${\cal R}_{-1}$ are
all equivalent (they can be seen as the result of the multiplication of a triplet by the singlets
$1$, $1'$, $1''$, respectively), in this 6D framework their difference is not irrelevant when building up
local interactions covariant under $A_4$ .

Generalizing what done above, a local invariant $J^{(N)}$ of degree $N$, built out of
$M$ brane multiplets $a^{(I)}$ $(I=1,...,M)$ transforming as ${\cal R}_{r_I}$ is
given by:
\be
J^{(N)}=\sum_{i=1}^{4} (a^{(1)}_i)^{n_1} \cdot\cdot\cdot  
(a^{(M)}_i)^{n_M} \delta_i~~~,
\ee
where $\sum_{I=1}^M n_I=N$ and $\sum_{I=1}^M r_I=0$ (mod 3).

\subsection{Bulk and brane fields}

Here we consider the coupling between a bulk multiplet ${\bf B}(z)=({\bf B}_1(z),{\bf B}_2(z),{\bf B}_3(z))$, transforming
as a triplet of $A_4$, and a brane multiplet $a=(a_1,a_2,a_3,a_4)$, transforming as ${\cal R}_0$
under $A_4$. The dependence on the 4D space-time coordinates $x$ is not made explicit in our notation.
For the time being, we assume that the three components ${\bf B}_I(z)$ $(I=1,2,3)$
are scalars in 6D.
The transformations of ${\bf B}$ under $A_4$ are specified by:
\be
\begin{array}{cll}
{\cal S}:& {\bf B}'(z_S)=S_3 {\bf B}(z)&z_S=z+\frac{1}{2}\\
{\cal T}:& {\bf B}'(z_T)=T_3 {\bf B}(z)&z_T=\omega z
\end{array}~~~.
\label{stbulk}
\ee  
 We write the most general local term bilinear in $a$ and ${\bf B}$ as:
\be
J=\sum_{i K} \alpha_{iK} a_i {\bf B}_K(z) \delta_i~~~,
\label{bubr}
\ee
where $\alpha_{iK}$ is a four by three matrix of constant coefficients.
It is not difficult to see that, in order to have $J$ invariant under $A_4$,
we should choose
\be
\alpha_{iK}=\frac{1}{2}
\left(
\begin{array}{ccc}
-1& +1& +1\\
+1& -1& +1\\
+1& +1& -1\\ 
-1& -1&-1
\end{array}
\right)~~~,
\label{inv}
\ee
up to an overall constant. If the brane multiplet $a$ is a ${\cal R}_0$ triplet under $A_4$, 
parametrized as in eq. (\ref{parab}), by choosing $\alpha_{iK}$ as in (\ref{inv}),
after integration over $z$ we get:
\bea
J&=&
\frac{1}{4}(-v_1+v_2+v_3)(-{\bf B}_1(z_1)+{\bf B}_2(z_1)+{\bf B}_3(z_1))+\\\nonumber
&+&\frac{1}{4}(+v_1-v_2+v_3)(+{\bf B}_1(z_2)-{\bf B}_2(z_2)+{\bf B}_3(z_2))\\\nonumber
&+&\frac{1}{4}(+v_1+v_2-v_3)(+{\bf B}_1(z_3)+{\bf B}_2(z_3)-{\bf B}_3(z_3))\\\nonumber
&+&\frac{1}{4}(+v_1+v_2+v_3)(+{\bf B}_1(z_4)+{\bf B}_2(z_4)+{\bf B}_3(z_4))
\eea
If the triplet ${\bf B}(z)$ acquires a constant VEV $\langle {\bf B}(z) \rangle=({\bf B}_1,{\bf B}_2,{\bf B}_3)$, 
essentially the only case that will be relevant for the discussion in the next session,
then the invariant $J$ becomes
\be
J=v_1 {\bf B}_1+v_2 {\bf B}_2+v_3 {\bf B}_3~~~.
\ee
Similarly, by requiring that $J'$ given by 
\be
J'=\sum_{i K} \alpha'_{iK} a_i {\bf B}_K(z) \delta_i~~~,
\label{bubrprime}
\ee
transforms as a $1'$, we find that the matrix
$\alpha'_{iK}$ should be given by
\be
\alpha'_{iK}=\frac{1}{2}
\left(
\begin{array}{ccc}
-1& +\omega& +\omega^2\\
+1& -\omega& +\omega^2\\
+1& +\omega& -\omega^2\\
-1& -\omega& -\omega^2\\
\end{array}
\right)~~~.
\label{dprime}
\ee
In this case, after integration over $z$ and after substitution of the triplet ${\bf B}(z)$
with its constant VEV,  the quantity $J$ of eq. (\ref{bubr}) becomes
\be
J=v_1 {\bf B}_1+\omega v_2 {\bf B}_2+\omega^2 v_3 {\bf B}_3~~~.
\ee
Finally, the singlet $1''$ is obtained from $J'$, by substituting
$\alpha'_{iK}$ with its complex conjugate $\alpha''_{iK}$.
\vskip 1.0cm 

\section{Orbifold realization of the $A_4$ model}

Let's start by recalling the basic formulae for the baseline $A_4$ model
for lepton masses and mixings in 4D with supersymmetry \cite{Mod}. 
The full superpotential of the model is
\be
w=w_l+w_d
\ee 
where $w_l$ is the term responsible for the Yukawa interactions 
in the lepton sector and $w_d$ is the term responsible for the vacuum alignment.
We now detail the structure of both in succession. The term $w_l$ is given
by
\be
w_l=y_e e^c (\varphi_T l)+y_\mu \mu^c (\varphi_T l)''+
y_\tau \tau^c (\varphi_T l)'+ (x_a\xi+\tilde{x}_a\tilde{\xi}) (ll)
+x_b (\varphi_S ll)+h.c.+...
\label{wlplus}
\ee
To keep our formulae compact, we omit to write the Higgs fields
$h_{u,d}$ and the cut-off scale $\Lambda$. For instance 
$y_e e^c (\varphi_T l)$ stands for $y_e e^c (\varphi_T l) h_d/\Lambda$,
$x_a\xi (ll)$ stands for $x_a\xi (l h_u l h_u)/\Lambda^2$ and so on.
The superpotential  $w_l$ contains the lowest order operators
in an expansion in powers of $1/\Lambda$. Dots stand for higher
dimensional operators. 
The ``driving'' term $w_d$ reads:
\bea
w_d&=&M (\varphi_0^T \varphi_T)+ g (\varphi_0^T \varphi_T\varphi_T)\nn\\
&+&g_1 (\varphi_0^S \varphi_S\varphi_S)+
g_2 \tilde{\xi} (\varphi_0^S \varphi_S)+
g_3 \xi_0 (\varphi_S\varphi_S)+
g_4 \xi_0 \xi^2+
g_5 \xi_0 \xi \tilde{\xi}+
g_6 \xi_0 \tilde{\xi}^2~~~,
\label{wd}
\eea
where $\varphi_0^T$, $\varphi_0^S$
and $\xi_0$ are driving fields that allow to build a 
non-trivial scalar potential in the symmetry breaking sector. 
The superpotential $w$ 
is invariant not only with respect to the gauge symmetry 
SU(2)$\times$ U(1) and the flavour symmetry $A_4$,
but also under a discrete $Z_3$ symmetry and a continuous U(1)$_R$ 
symmetry under which the fields 
transform as shown in the following table.
\\[0.2cm]
\begin{table}[h]
\begin{center}
\begin{tabular}{|c||c|c|c|c||c|c|c|c|c||c|c|c|}
\hline
{\tt Field}& l & $e^c$ & $\mu^c$ & $\tau^c$ & $h_{u,d}$ & 
$\varphi_T$ & $\varphi_S$ & $\xi$ & $\tilde{\xi}$ & $\varphi_0^T$ & $\varphi_0^S$ & $\xi_0$\\
\hline
$A_4$ & $3$ & $1$ & $1'$ & $1''$ & $1$ & 
$3$ & $3$ & $1$ & $1$ & $3$ & $3$ & $1$\\
\hline
$Z_3$ & $\omega$ & $\omega^2$ & $\omega^2$ & $\omega^2$ & $1$ &
$1$ & $\omega$ & $\omega$ & $\omega$ & $1$ & $\omega$ & $\omega$\\
\hline
$U(1)_R$ & $1$ & $1$ & $1$ & $1$ & $0$ & 
$0$ & $0$ & $0$ & $0$ & $2$ & $2$ & $2$\\
\hline
\end{tabular}
\end{center}
\caption{Fields and their transformation properties under $A_4$, $Z_3$ and $U(1)_R$.}
\end{table}
\vspace{0.2cm}

We now show how this model can be derived from the 6D field theory 
with orbifolding. We start from an $N=1$ chiral supersymmetric 6D field 
theory, corresponding to $N=2$ SUSY in the 4D language. Such an extended SUSY
is broken down to $N=1$ SUSY by the $Z_2$ parity in the usual way.
The lagrangian of the theory is the sum of a bulk term, depending on
bulk fields and invariant under $N=2$ SUSY, plus boundary terms
localized at the four fixed points and invariant under the less restrictive
$N=1$ SUSY. Moreover at the fixed points we are allowed to localize
brane $N=1$ multiplets. In particular we choose as brane fields
the gauge bosons of the SM gauge group, the SM fermions and two Higgs
doublets $h_u$ and $h_d$, together with their $N=1$ superpartners. The remaining fields,
namely the flavons and the driving fields
are introduced as bulk hypermultiplets. 
In this way we avoid 6D gauge anomalies. Due to the orbifolding,
out of the two $N=1$ chiral supermultiplets contained in the generic 
bulk hypermultiplet only one possesses a zero mode.
Here we are interested in the brane interactions of this particular 
multiplet, and we will use for it the $N=1$ notation.

The dictionary between the 4D realization, specified by the superpotential $w_l$ and
the present 6D version, is given in table 2.
We have denoted by $l_i$ the lepton doublet supermultiplets,
which are $A_4$-triplet brane fields parametrized as in eq. (\ref{parab}):
\\[0.2cm]
\begin{table}[h]
\begin{center}
\begin{tabular}{|c|c|}
\hline
&\\
{\tt 4D}& {\tt 6D} \\
&\\
\hline
&\\
$\xi (ll)$& $\sum_{i=1}^4 l_i l_i \mbox{\boldmath $\xi$}(z) \delta_i$\\
&\\
\hline
&\\
$(\varphi_S l l)$& $\sum_{i=1}^4 l_i l_i \alpha_{iK} {\mbox{\boldmath $\varphi$}_S}_K(z) \delta_i$\\
&\\
\hline
&\\
$e^c (\varphi_T l)$& $\sum_{i=1}^4 e^c l_i \alpha_{iK} {\mbox{\boldmath $\varphi$}_T}_K(z) \delta_i$\\
&\\
\hline
&\\
$\mu^c (\varphi_T l)''$& $\sum_{i=1}^4 \mu^c l_i \alpha''_{iK}  {\mbox{\boldmath $\varphi$}_T}_K(z)\delta_i$\\
&\\
\hline
&\\
$\tau^c (\varphi_T l)'$& $\sum_{i=1}^4 \tau^c l_i \alpha'_{iK} {\mbox{\boldmath $\varphi$}_T}_K(z)\delta_i$\\
&\\
\hline
\end{tabular}
\end{center}
\caption{Realization of 4D superpotential terms for $w_l$ in terms of local 6D
$A_4$ invariants. The 4D terms are obtained from the 6D ones by integrating over
the complex coordinate $z$ and by assuming a constant background value for the bulk multiplets
$\langle \mbox{\boldmath $\varphi$}_{S,T}(z)\rangle=\langle\varphi_{S,T}\rangle/\sqrt{V}$, 
$\langle \mbox{\boldmath $\xi$}(z)\rangle=\langle \xi\rangle/\sqrt{V}$.}
\end{table}
\vspace{0.2cm}
\be
l=\frac{1}{2}
\left(
\begin{array}{c}
-l_e+l_\mu+l_\tau\\
+l_e-l_\mu+l_\tau\\
+l_e+l_\mu-l_\tau\\
-l_e-l_\mu-l_\tau
\end{array}
\right)~~~.
\label{paral}
\ee
The charged leptons $e^c$, $\mu^c$ and $\tau^c$ are brane fields, having
the same value at each fixed point.
As anticipated, the flavon fields \mbox{\boldmath $\varphi$}$_S(z)$, \mbox{\boldmath $\varphi$}$_T(z)$ and \mbox{\boldmath $\xi$}$(z)$
are bulk fields, depending on the extra coordinate $z$. In particular \mbox{\boldmath $\varphi$}$_S(z)$
and \mbox{\boldmath $\varphi$}$_T(z)$ are $A_4$ triplets, transforming as in eq. (\ref{stbulk}),
while \mbox{\boldmath $\xi$}$(z)$ is an $A_4$ invariant: \mbox{\boldmath $\xi$}$'(z+1/2)={\bf\xi}(z)$ and \mbox{\boldmath $\xi$}$'(\omega z)={\bf\xi}(z)$.
Each 4D superpotential term is reproduced, up to an overall constant, from the
corresponding 6D term of the dictionary by integrating over the complex coordinate
$z$ and by assuming a constant, that is $z$-independent, background value
for the bulk supermultiplets \mbox{\boldmath $\varphi$}$_S(z)$, \mbox{\boldmath $\varphi$}$_T(z)$ and \mbox{\boldmath $\xi$}$(z)$.
This last requirement is justified by the fact that we only need to discuss the
expansion of $w$ around the VEVs of the flavon fields. Barring a peculiar 
behaviour of such VEVs, we will look for minima of the scalar potential that
do not depend on $z$ and in our final expressions the bulk fields will be
replaced by their constant VEVs.
In this way the superpotential $w_l$ is completely reproduced.

To correctly establish the relation between the 6D superpotential and the 4D one
we should also pay attention to the overall normalization
of $w_l$. The 6D superpotential $w_l$ is linear in the bulk fields having
mass dimension two and therefore carries an extra factor $1/\Lambda$
with respect to the 4D superpotential. Moreover, the VEV of the generic bulk field
${\bf B}$ can be parametrized as $\langle B \rangle/\sqrt{V}$ where
$\langle B \rangle$ is the VEV of the zero mode, of mass dimension one, and
$V$ is the volume of the extra compact space. Therefore, after spontaneous
breaking of the $A_4$ symmetry, each bulk field ${\bf B}$ enters the superpotential
in the dimensionless combination $\langle B\rangle/(\Lambda^2\sqrt{V})$. Higher dimensional operators
are suppressed by extra powers of this combination. To avoid large corrections to 
the HPS mixing scheme, such a combination is required to be at most of order
$\lambda^2$, $\lambda\approx 0.22$ being the Cabibbo angle.
This is of no concern for the lepton sector of the theory, but it can be a potential 
problem for the extension of the $A_4$ model, both in its 4D and 6D realizations, to the quark sector. 
Indeed we expect that the mass of the top quark
arises from an unsuppressed renormalizable operator, whereas a naive extension of
the $A_4$ assignment to the quark sector of our 6D model would lead to a top
mass depleted by an overall factor $\langle \varphi_T\rangle/(\Lambda^2 \sqrt{V})$ (with respect, say, to the $W$ mass), which
as we have seen is expected to be of order $\lambda^2$.

Finally we need a similar dictionary for the driving part of the superpotential.
It is easy to see that each 4D term in $w_d$ can be reproduced starting from
a corresponding 6D term, by assuming constant field configurations and by 
integrating over the coordinate $z$. The new feature when analyzing $w_d$ is that
in general there is no one-to-one correspondence between 4D and 6D terms
as was the case for $w_l$ because the number of local 6D invariants we can build from
bulk fields is larger than the number of 4D invariants we have in $w_d$.
This is not an obstacle in deriving the 4D theory. Since we are interested
in constant field configurations of the flavon and driving fields,
after integration over $z$ our 6D driving superpotential will indeed
give rise to the most general set of $A_4$ invariants in 4D. The result is nothing but
the superpotential $w_d$ given in eq. (\ref{wd}). At this point the
discussion of the vacuum alignment proceeds exactly as in the 4D case,
detailed in ref. \cite{Mod}.

The scalar potential is minimum at:
\be
\begin{array}{cclr}
\langle \mbox{\boldmath $\varphi$}_T \rangle&=&\dd\frac{1}{\sqrt{V}}(v_T,v_T,v_T)&\dd\frac{v_T}{\Lambda\sqrt{V}}=-\dd\frac{M}{g}\\
\langle \mbox{\boldmath $\varphi$}_S \rangle&=&\dd\frac{1}{\sqrt{V}}(v_S,0,0)&v_S^2=-\dd\frac{g_4}{g_3} u^2\\
\langle \mbox{\boldmath $\xi$} \rangle&=&\dd\frac{1}{\sqrt{V}} u& u~~ {\tt undetermined}\\
\langle \mbox{\boldmath $\tilde{\xi}$} \rangle&=&0&~~~.
\end{array}
\ee
At the leading order of the $1/\Lambda$ expansion,
the mass matrix $m_l$ for charged leptons is given by:
\be
m_l=v_d\frac{v_T}{\Lambda^2\sqrt{V}}\left(
\begin{array}{ccc}
y_e& y_e& y_e\\
y_\mu& y_\mu \omega& y_\mu \omega^2\\
y_\tau& y_\tau \omega^2& y_\tau \omega
\end{array}
\right)~~~,
\label{mch}
\ee
and charged fermion masses are given by:
\be
m_e=\sqrt{3} y_e v_d \frac{v_T}{\Lambda^2\sqrt{V}} ~~~,~~~~~~~
m_\mu=\sqrt{3} y_\mu v_d \frac{v_T}{\Lambda^2\sqrt{V}}~~~,~~~~~~~
m_\tau=\sqrt{3} y_\tau v_d \frac{v_T}{\Lambda^2\sqrt{V}}~~~.
\label{chmasses}
\ee
We can easily obtain a natural hierarchy among $m_e$, $m_\mu$ and
$m_\tau$ by introducing an additional U(1)$_F$ flavour symmetry under
which only the right-handed lepton sector is charged.
In the flavour basis the neutrino mass matrix reads :
\be
m_\nu=\frac{v_u^2}{\Lambda}\left(
\begin{array}{ccc}
a+2 d/3& -d/3& -d/3\\
-d/3& 2d/3& a-d/3\\
-d/3& a-d/3& 2 d/3
\end{array}
\right)~~~,
\label{mnu0}
\ee
where
\be
a\equiv x_a\frac{u}{\Lambda^2 \sqrt{V}}~~~,~~~~~~~d\equiv x_d\frac{v_S}{\Lambda^2 \sqrt{V}}~~~,
\label{ad}
\ee
and is diagonalized by the transformation:
\be
U^T m_\nu U =\frac{v_u^2}{\Lambda}{\tt diag}(a+d,a,-a+d)~~~,
\ee
with
\be
U=\left(
\begin{array}{ccc}
\sqrt{2/3}& 1/\sqrt{3}& 0\\
-1/\sqrt{6}& 1/\sqrt{3}& -1/\sqrt{2}\\
-1/\sqrt{6}& 1/\sqrt{3}& +1/\sqrt{2}
\end{array}
\right)~~~.
\ee
For the neutrino masses we obtain:
\bea
|m_1|^2&=&\left[-r+\frac{1}{8\cos^2\Delta(1-2r)}\right]
\Delta m^2_{atm}\nn\\
|m_2|^2&=&\frac{1}{8\cos^2\Delta(1-2r)}
\Delta m^2_{atm}\nn\\
|m_3|^2&=&\left[1-r+\frac{1}{8\cos^2\Delta(1-2r)}\right]\Delta m^2_{atm}~~~,
\label{lospe}
\eea
where $r\equiv \Delta m^2_{sol}/\Delta m^2_{atm}
\equiv (|m_2|^2-|m_1|^2)/|m_3|^2-|m_1|^2)$,
$\Delta m^2_{atm}\equiv|m_3|^2-|m_1|^2$ 
and $\Delta$ is the phase difference between
the complex numbers $a$ and $d$. For  $\cos\Delta=-1$, we have a neutrino spectrum close to hierarchical:
\be
|m_3|\approx 0.053~~{\rm eV}~~~,~~~~~~~
|m_1|\approx |m_2|\approx 0.017~~{\rm eV}~~~.
\ee
\begin{figure}[h!]
\centerline{\psfig{file=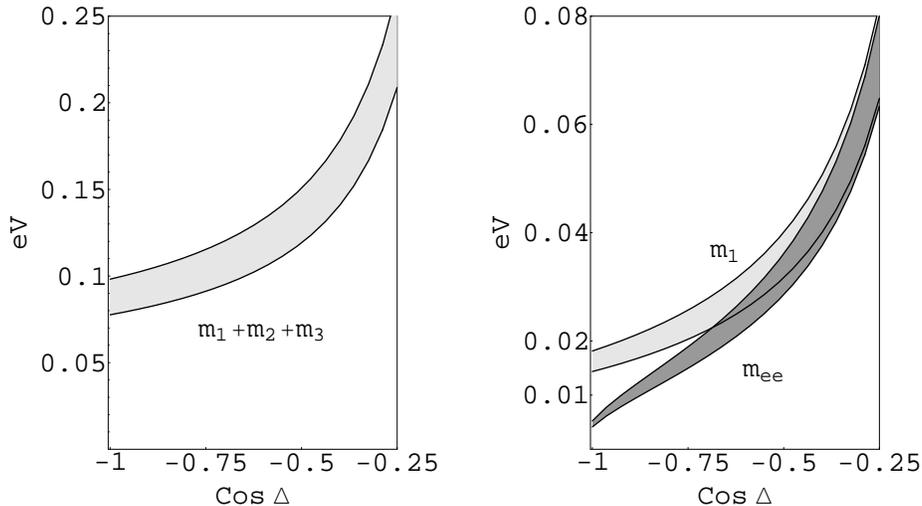,width=0.8\textwidth}}
\caption{On the left panel, sum of neutrino masses versus $\cos\Delta$, the phase difference between $a$ and $b$. On the right panel, the lightest neutrino mass, $m_1$ and the mass combination $m_{ee}$ versus $\cos\Delta$.
To evaluate the masses, the parameters $\vert a\vert$ and $\vert b\vert$
have been expressed in terms of $r\equiv \Delta m^2_{sol}/\Delta m^2_{atm}
\equiv (|m_2|^2-|m_1|^2)/|m_3|^2-|m_1|^2)$ and
$\Delta m^2_{atm}\equiv|m_3|^2-|m_1|^2$. The bands have been obtained by 
varying $\Delta m^2_{atm}$ in its 3$\sigma$ experimental range, 0.0020 eV $\div$ 0.0032 eV. There is a negligible sensitivity to the variations of $r$
within its current 3 $\sigma$ experimental range, and we have
realized the plots by choosing $r=0.03$.}
\end{figure} 
In this case the sum of neutrino masses is about $0.087$ eV.
If $\cos\Delta$ is accidentally small, the neutrino spectrum becomes
degenerate. The value of $|m_{ee}|$, the parameter characterizing the 
violation of total lepton number in neutrinoless double beta decay,
is given by:
\be
|m_{ee}|^2=\left[-\frac{1+4 r}{9}+\frac{1}{8\cos^2\Delta(1-2r)}\right]
\Delta m^2_{atm}~~~.
\ee
For $\cos\Delta=-1$ we get $|m_{ee}|\approx 0.005$ eV, at the upper edge of
the range allowed for normal hierarchy, but unfortunately too small
to be detected in a near future.
Independently from the value of the unknown phase $\Delta$
we get the relation:
\be
|m_3|^2=|m_{ee}|^2+\frac{10}{9}\Delta m^2_{atm}\left(1-\frac{r}{2}\right)~~~,
\ee
which is a prediction of our model. In Fig. 2 we have plotted the neutrino masses predicted by the model.

In summary, we have obtained the baseline 4D $A_4$ model starting from a 6D
realization, where all SM supermultiplets live at the fixed points of a $T^2/Z_2$
orbifold and the flavon and driving fields live in the bulk.
 
\section{Conclusion}

We have shown that extra dimensional theories with orbifolding provide a natural framework to interpret 
flavour symmetries as  discrete permutational symmetries among fixed point branes. In particular, starting 
from a 6D theory, we have discussed an orbifolding with 4 fixed points leading to the $A_4$ flavour 
symmetry. In this picture $A_4$ together with 4D translations and 4D proper Lorentz transformations 
represents the subgroup of 6D space-time symmetry which 
is left unbroken in the theory after orbifolding and after locating the SM particles on the fixed point 
branes. Note that $A_4$ and not the full permutation group $S_4$ is the residual symmetry group because 
only even permutations can be seen as the result of a rigid space rotation. Each brane field, either a 
triplet or a singlet, has components on all of the four fixed points (in particular all components are 
equal for a singlet) but the interactions are local, i.e. all vertices involve products of field 
components at the same space-time point. This approach suggests a deep relation between flavour symmetry 
in 4D and  space-time symmetry in extra dimensions. We have also demonstrated that a SUSY model of neutrino 
tri-bimaximal mixing based on $A_4$, which we have formulated in a recent work \cite{Mod}, can be 
directly reinterpreted in the orbifolding approach.

\section*{Acknowledgements}
We recognize that this work has been partly supported by the European Commission under contract MRTN-CT-2004-503369.


\begin{thebibliography}{99}

\bibitem{data}
T.~Schwetz,
  Phys.\ Scripta {\bf T127} (2006) 1
  [arXiv:hep-ph/0606060];
  M.~Maltoni, T.~Schwetz, M.~A.~Tortola and J.~W.~F.~Valle,
  New J.\ Phys.\  {\bf 6} (2004) 122, see arXiv:hep-ph/0405172 v5;
 A.~Strumia and F.~Vissani,
  Nucl.\ Phys.\ B {\bf 726}, 294 (2005)
  [arXiv:hep-ph/0503246];
G.~L.~Fogli {\it et al.},
  arXiv:hep-ph/0608060.

\bibitem{hps}
P.~F.~Harrison, D.~H.~Perkins and W.~G.~Scott,
Phys.\ Lett.\ B {\bf 530} (2002) 167
[arXiv:hep-ph/0202074];
P.~F.~Harrison and W.~G.~Scott,
Phys.\ Lett.\ B {\bf 535} (2002) 163
[arXiv:hep-ph/0203209];
Z.~z.~Xing,
Phys.\ Lett.\ B {\bf 533} (2002) 85
[arXiv:hep-ph/0204049];
P.~F.~Harrison and W.~G.~Scott,
Phys.\ Lett.\ B {\bf 547} (2002) 219
[arXiv:hep-ph/0210197];
P.~F.~Harrison and W.~G.~Scott,
Phys.\ Lett.\ B {\bf 557} (2003) 76
[arXiv:hep-ph/0302025];
P.~F.~Harrison and W.~G.~Scott,
arXiv:hep-ph/0402006;
P.~F.~Harrison and W.~G.~Scott,
arXiv:hep-ph/0403278.

\bibitem{ma1}
E.~Ma and G.~Rajasekaran,
  Phys.\ Rev.\ D {\bf 64} (2001) 113012
  [arXiv:hep-ph/0106291].
 
 \bibitem{ma2}
 E.~Ma,
  Mod.\ Phys.\ Lett.\ A {\bf 17} (2002) 627
  [arXiv:hep-ph/0203238];
K.~S.~Babu, E.~Ma and J.~W.~F.~Valle,
Phys.\ Lett.\ B {\bf 552} (2003) 207
[arXiv:hep-ph/0206292];
M.~Hirsch, J.~C.~Romao, S.~Skadhauge, J.~W.~F.~Valle and A.~Villanova del Moral
[arXiv:hep-ph/0312244];
[arXiv:hep-ph/0312265];
E.~Ma,
  Phys.\ Rev.\ D {\bf 70} (2004) 031901
  [arXiv:hep-ph/0404199];
E.~Ma
  arXiv:hep-ph/0409075;
E.~Ma,
  New J.\ Phys.\  {\bf 6} (2004) 104;
S.~L.~Chen, M.~Frigerio and E.~Ma,
  Nucl.\ Phys.\ B {\bf 724} (2005) 423
  [arXiv:hep-ph/0504181];
E.~Ma,
  Phys.\ Rev.\ D {\bf 72} (2005) 037301
  [arXiv:hep-ph/0505209];
K.~S.~Babu and X.~G.~He,
  arXiv:hep-ph/0507217;
A.~Zee,
  Phys.\ Lett.\ B {\bf 630} (2005) 58
  [arXiv:hep-ph/0508278];
E.~Ma,
  Mod.\ Phys.\ Lett.\ A {\bf 20} (2005) 2601
  [arXiv:hep-ph/0508099];
E.~Ma,
  arXiv:hep-ph/0511133;
S.~K.~Kang, Z.~z.~Xing and S.~Zhou,
  Phys.\ Rev.\ D {\bf 73}, 013001 (2006)
  [arXiv:hep-ph/0511157];
  X.~G.~He, Y.~Y.~Keum and R.~R.~Volkas,
  JHEP {\bf 0604} (2006) 039
  [arXiv:hep-ph/0601001];
  B.~Adhikary, B.~Brahmachari, A.~Ghosal, E.~Ma and M.~K.~Parida,
  Phys.\ Lett.\ B {\bf 638} (2006) 345
  [arXiv:hep-ph/0603059];
    E.~Ma,
  arXiv:hep-ph/0607190;
L.~Lavoura and H.~Kuhbock,
  arXiv:hep-ph/0610050.

\bibitem{continuous}
S.~F.~King,
  JHEP {\bf 0508} (2005) 105
  [arXiv:hep-ph/0506297];
I.~de Medeiros Varzielas and G.~G.~Ross,
  Nucl.\ Phys.\ B {\bf 733} (2006) 31
  [arXiv:hep-ph/0507176];
S.~F.~King and M.~Malinsky,
  arXiv:hep-ph/0608021.

\bibitem{others}
For others approaches to the tri-bimaximal mixing see:
J.~Matias and C.~P.~Burgess,
  JHEP {\bf 0509} (2005) 052
  [arXiv:hep-ph/0508156];
S.~Luo and Z.~z.~Xing,
  arXiv:hep-ph/0509065;
W.~Grimus and L.~Lavoura,
  arXiv:hep-ph/0509239;
F.~Caravaglios and S.~Morisi,
  arXiv:hep-ph/0510321;
I  .~de Medeiros Varzielas, S.~F.~King and G.~G.~Ross,
  [arXiv:hep-ph/0512313];
  I.~de Medeiros Varzielas, S.~F.~King and G.~G.~Ross,
  [arXiv:hep-ph/0607045];
C.~Hagedorn, M.~Lindner and R.~N.~Mohapatra,
  JHEP {\bf 0606} (2006) 042
  [arXiv:hep-ph/0602244];
P.~Kovtun and A.~Zee,
  Phys.\ Lett.\ B {\bf 640} (2006) 37
  [arXiv:hep-ph/0604169];
R.~N.~Mohapatra, S.~Nasri and H.~B.~Yu,
  Phys.\ Lett.\ B {\bf 639} (2006) 318
  [arXiv:hep-ph/0605020];
Z.~z.~Xing, H.~Zhang and S.~Zhou,
  Phys.\ Lett.\ B {\bf 641} (2006) 189
  [arXiv:hep-ph/0607091].


\bibitem{OurTriBi}
G.~Altarelli and F.~Feruglio,
  Nucl.\ Phys.\ B {\bf 720} (2005) 64
  [arXiv:hep-ph/0504165].
 
 \bibitem{Mod}
G.~Altarelli and F.~Feruglio,
  Nucl.\ Phys.\ B {\bf 741} (2006) 215
  [arXiv:hep-ph/0512103].

\bibitem{gpr}
See for instance A.~Giveon, M.~Porrati and E.~Rabinovici,
  Phys.\ Rept.\  {\bf 244}, 77 (1994)
  [arXiv:hep-th/9401139];
 For more specific examples see:
 W.~Lerche, D.~Lust and N.~P.~Warner,
  Phys.\ Lett.\ B {\bf 231}, 417 (1989);
  E.~J.~Chun, J.~Mas, J.~Lauer and H.~P.~Nilles,
  Phys.\ Lett.\ B {\bf 233}, 141 (1989);
S.~Ferrara, D.~Lust and S.~Theisen,
  Phys.\ Lett.\ B {\bf 233}, 147 (1989).
\bibitem{watyan}
T.~Watari and T.~Yanagida,
  Phys.\ Lett.\ B {\bf 532}, 252 (2002)
  [arXiv:hep-ph/0201086] and 
  Phys.\ Lett.\ B {\bf 544}, 167 (2002)
  [arXiv:hep-ph/0205090].
\end{thebibliography}
\end{document}